\documentclass[letter,twocolumn]{jpsj2}

\usepackage{graphicx}
\def\runtitle{%
Theory of magnetotunneling spectroscopy in 
spin triplet $p$-wave superconductors }
\def\runauthor{%
Yukio {\sc Tanaka}, 
Yasunari {\sc Tanuma}, 
Kazuhiko {\sc Kuroki},  
and Satoshi {\sc Kashiwaya}}
\title{%
Theory of magnetotunneling spectroscopy in 
spin triplet $p$-wave superconductors}

\author{%
Yukio {\sc Tanaka}$^{1,2}$, 
Yasunari {\sc Tanuma}$^{3}$,
Kazuhiko {\sc Kuroki}$^{4}$, 
and Satoshi {\sc Kashiwaya}$^{2,5}$
}

\inst{%
$^{1}$ 
Department of Applied Physics,
Nagoya University, Nagoya 464-8063  Japan 
\\
$^{2}$ Crest, Japan Science and Technology Corporation (JST), 
Nagoya 464-8063 Japan
\\
$^{3}$ Graduate School of Natural Science and Technology, 
Okayama University,  Okayama 700-8530, Japan
\\
$^{4}$ 
Department of Applied Physics and Chemistry,  
The University of Electro-Communications Chofu Tokyo 
182-8585, Japan
\\
$^{5}$ Natural Institute of Advanced Industrial Science and 
Technology, Tsukuba, 305-8568,  Japan
}

\recdate{\today}

\abst{%
We study the influence of a magnetic field $H$ on the
zero-bias conductance peak (ZBCP) due to 
zero-energy Andreev bound state (ZES) in 
normal metal / unconventional superconductor.
For $p$-wave junctions, 
ZBCP does not split into two 
by $H$ even for sufficiently 
low transparent junctions,
where ZBCP clearly splits for $d$-wave. 
This unique property originates from the fact that 
for $p$-wave superconductors, perpendicularly injected 
quasiparticle form ZES, which contribute
most dominantly on the tunneling conductance. 
In addition, we show that for 
$p_{x}$+i$p_{y}$-wave superconductor junctions, 
the height of ZBCP is sensitive to $H$ due to the 
formation of broken time reversal symmetry state.
We propose that tunneling spectroscopy 
in the presence of magnetic field,
$i.e.$, $magnetotunneling$, is an promising method
to determine the pairing symmetry of 
unconventional  superconductors.
\par
}
\kword{%
triplet superconductor, tunneling conductance, 
Doppler shift, chirality }
\begin{document}
%\sloppy
\maketitle
%=========INTRODUCTION===========================================
%\section{Introduction}
%
%-----------------------------------------------------------
Nowadays, 
the formation of zero-energy Andreev bound states (ZES)
at surfaces \cite{Buch,Hu} or interfaces \cite{Kasi00} of 
unconventional superconductors is receiving increasing attention.
This state, which is created by injected and reflected quasiparticle's 
feeling different signs of the pair potential,
can play an important role in determining the 
pairing symmetry of anisotropic superconductors.
The detection of the ZES is reflected as
an observation of a zero-bias conductance peak (ZBCP)
in tunneling conductance,
which is generally considered as a clear signature of
anisotropic pairing \cite{Tanaka95,Kashi96,Honer98,Boss00,Boss01}.
The ZBCP is observed
in various superconductors having anisotropic pairing symmetry, 
such as the cuprates 
\cite{Kashi95,Alff,Wei,Wang,Iguchi}.
$\mbox{Sr}_2\mbox{RuO}_{4}$ \cite{Laube,Mao}, 
and $\mbox{UBe}_{13}$ \cite{Walti}. 
Possibility of ZBCP has also been theoretically predicted for 
organic superconductors $\mbox{(TMTSF)}_{2}X$
very recently \cite{Sengupta,Tanu02,TAU2002}.
The existence of the ABS has crucial influences on many 
transport phenomena in unconventional 
superconductor junctions 
\cite{Kasi00,K99,T1,T2,Hirai,Barash,Asano}.  
\par
Since the existence of ZES is a 
universal phenomena expected for unconventional superconductors 
having pair potential that changes sign on the Fermi surface, 
difficulty may arise in determining the pairing symmetry
only from conventional tunneling spectroscopy.  
%
%unless the detailed study for various orientation of the 
%well controlled interface is realized. 
% 
%However, at this stage, it is not easy to prepare 
%well controlled interface  with various  orientations 
%\cite{Iguchi}. 
\par
In order to overcome this difficulty,
we require some $in$ $situ$ way 
of probing the symmetry from the tunneling spectroscopy. 
Here, we will show that a promising way is to use a 
tunneling spectroscopy in the presence of a 
magnetic field. 
For $d_{x^{2}-y^{2}}$-wave junction, 
it has been shown that 
screening currents shift the ZES spectrum (Doppler shift)
and lead to a splitting of ZBCP 
\cite{Fogel,Covington,Aprili,Krupke}.
\par
By contrast, we show in the present paper that for $p$-wave cases, 
ZBCP does not split into two in the presence of a magnetic field 
since the most dominant contribution to tunneling conductance 
is given by perpendicular injection,
where the energy shift of the quasiparticles
does not occur at all
because the component of the Fermi velocity 
parallel to the interface is zero
for a cylindrical Fermi surface.
We also show phenomenologically the absence of Doppler shift
for more general shapes of the Fermi surface, 
assuming inversion symmetry of crystal,
where the component of the Fermi velocity parallel to the interface 
always have the same magnitude but different signs between 
perpendicularly injected and reflected quasiparticle states.
Finally, we show that for a $p_{x}$+i$p_{y}$-wave superconductor,
the magnitude of ZBCP is enhanced or suppressed 
depending on which way ($+z$ or $-z$ direction)
to apply the magnetic field.  
We propose that this dependence on the magnetic field 
can be used in detecting the broken time 
reversal symmetry superconducting state.
\par
%
%=======================================================
%\section{Formulation}
%=======================================================
We calculate tunneling conductance in 
normal metal / unconventional superconductor 
junctions by solving  
Bogoliubov-de Gennes (BdG) equation 
using quasiclassical approximation as in our previous theory
\cite{Kasi00,Tana02}. 
As regards the triplet pairing cases, 
we assume that Cooper pairs are formed by electrons having 
antiparallel spins, but an extension to more general cases 
including parallel spin pairing or non-unitary cases 
\cite{Boss00,Boss01} is straightforward. 
%However, in the present paper, we only concentrate 
%on the odd parity of the triplet pairing as a function of 
%$\vct{ k}$, it is sufficient to discuss 
%Cooper pair composed with electrons with opposite 
%spin as a simplest case. 
%
Now, we consider the case where a specularly
reflecting surface or interface run along the $y$-direction. 
The insulator located at the interface between normal metal and 
$d$-wave superconductor is expressed using $U(x)$ as 
\begin{equation}
U(x)= 
\left\{ \begin{array}{ll}
0 & x<-d_{i} \\
U_{0} & -d_{i}<x<0 \\
0 & x>0
\end{array}
\right. 
\end{equation}
where the width and height of the barrier are 
$d_{i}$ and $U_{0}$, respectively. 
The magnetic field is applied parallel to the $z$-axis, 
so that the vector potential can be 
chosen as $\mib{A}(\mib{r})=(0,A_{y}(x),0)$. 
The pair potential depends only on $x$
since the system is homogeneous along the $y$-direction.
Since we are now considering the situation 
where the coherence length of the pair potential $\xi$ is 
much smaller than the penetration depth of the magnetic field 
$\lambda$, we can ignore the spatial dependence of 
$A_{y}(x)$ in the following calculations. 
We assume $A_{y}(x)=A_{0}=-H \lambda$, where $H$ is the applied 
magnetic field. 
The normalized tunneling conductance 
$\sigma_{\rm T}(eV)$ under the bias voltage $V$
is given by \cite{Kasi00,Tanaka95,Tana02}
\begin{align}
\sigma_{\rm T}(E) &=
\frac{\displaystyle 
\int^{\pi/2}_{-\pi/2}d\theta 
\sigma_{\rm S}(\theta,E)}
{\displaystyle
\int^{\pi/2}_{-\pi/2}d\theta 
\sigma_{\rm N}(\theta)
\cos \theta}.
\\
\nonumber
\sigma_{\rm S}(\theta,E) &= 
\sigma_{\rm N}(\theta)
\sigma_{\rm R}(\theta,E)\cos \theta,
\end{align}
\[
\sigma_{\rm N}(\theta)
=\frac{4Z^{2} }
{(1 -Z^{2})^{2} 
\sinh^{2}(\lambda_{0} d_{i}) 
+ 4Z^{2} \cosh^{2}(\lambda_{0} d_{i})},
\]
\[
Z=\frac{\kappa\cos\theta}
{\sqrt{1-\kappa^{2}\cos^{2}\theta }},  \ 
\lambda_{0}=\sqrt{2mU_{0}/\hbar^{2}}, \ 
\kappa=k_{\rm F}/\lambda_{0},
\]
\begin{eqnarray}
\sigma_{\rm R}(\theta,E)
=\frac{1+\sigma_{\rm N}(\theta)|\Gamma_{+}|^{2}
+[\sigma_{\rm N}(\theta)-1]
|\Gamma_{+}|^{2}|\Gamma_{-}|^{2}}
{|1+[\sigma_{\rm N}(\theta)-1]
\Gamma_{+} \Gamma_{-}|^{2}}.
\end{eqnarray}
\begin{equation}
\Gamma_{\pm} = 
\left\{ \begin{array}{ll}
\frac{\displaystyle 
 \Delta_{0}f(\theta_{\pm})}
{\displaystyle 
\sqrt{\tilde{E}_{\pm}^{2} -
\mid \Delta_{0} f(\theta_{\pm}) \mid^{2} }
+ \tilde{E}_{\pm}  }, & \tilde{E}_{\pm}  > 0, \\
\frac{\displaystyle -\Delta_{0} f(\theta_{\pm})}
{\displaystyle \sqrt{\tilde{E}_{\pm}^{2} -
\mid \Delta_{0} f(\theta_{\pm}) \mid^{2} }^{*}
- \tilde{E}_{\pm}  }, & \tilde{E}_{\pm}  < 0, 
\end{array}
\right. 
\end{equation} 

\[
\tilde{E}_{\pm} =E + ev_{{\rm F}y}H\lambda 
=E + \frac{\Delta_{0}H g_{\pm}}{H_{0}},
\]
with the injected and reflected angles being 
$\theta_{+}=\theta$ and  
$\theta_{-}=\pi -\theta$, respectively, 
$H_{0}=\frac{\Delta_{0}}{e \lambda v_{\rm F}}
=\frac{\phi_{0}}{\xi_{0} \lambda \pi^{2}}$, 
$\xi_{0}=\hbar v_{\rm F}/(\pi \Delta_{0})$, 
$E=eV$, and $g_{\pm}= v_{{\rm F}y}/v_{\rm F}$, 
where $k_{\rm F}$ and $v_{\rm F}$ is the Fermi momentum
and the Fermi velocity, respectively, taken to be common 
in superconductor and normal metal for simplicity.
$\Delta_{0}$ is the 
magnitude of the maximum value of the 
pair potential and  
$\tilde{E}_{\pm}$ are the Doppler shifted energies of 
injected and reflected quasiparticle states.
In the above, we have assumed a spatially 
constant pair potential in the 
superconductor.  
As shown in our previous paper, 
if we concentrate on the qualitative feature 
of $\sigma_{\rm T}(E)$ at low voltage, this 
assumption is justified 
\cite{Kasi00,Tana02}.
\par
%=======================================================
%\section{Results}
%=======================================================
We take $\lambda_{0}d_{i}=3$ and $\kappa=0.5$, where 
the transmission coefficient perpendicular 
to the interface is about 0.02.
In the following, 
we will look at $\sigma_{\rm T}(eV)$
and $\sigma_{\rm S}(\theta,eV)$ 
for various cases.
\par
%%%%%%%%%%%%%%%%%%%%%%%%%%%%%%%%%%%%%%%%%%%
% d wave & p-wave total conductance
%%%%%%%%%%%%%%%%%%%%%%%%%%%%%%%%%%%%%
%--- figure 1 ---
\begin{figure}[htb]
\begin{center}
\vspace{2cm}
\includegraphics[scale=.4]{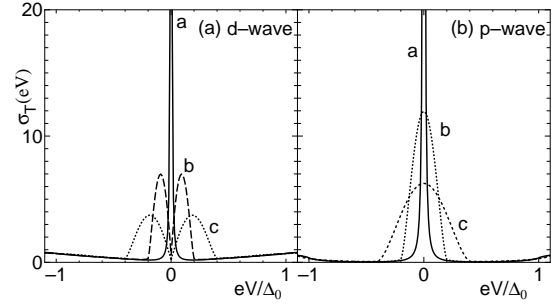}
\caption{
The normalized tunneling conductance $\sigma_{\rm T}(eV)$
in the (a) $d_{x^{2}-y^{2}}$-wave state 
with (110) oriented surface and (b) $p_{x}$-wave state with
(100) oriented surface. 
$\lambda_{0}d_{i}=3$ and $\kappa=0.5$. 
a: $H=0$,
b: $H=0.2H_{0}$, and 
c: $H=0.4H_{0}$.
\label{fig:01}}
\end{center}
\end{figure}
%----------------
We have calculated $\sigma_{\rm T}(eV)$
for a $d_{x^{2}-y^{2}}$-wave junction with 
(110) oriented surface and for a $p_{x}$-wave junction with 
(100) oriented surface, where $f(\theta)$ is expressed as 
$f(\theta)=\sin(2\theta)$ and $f(\theta)=\cos(\theta)$, respectively, 
as a prototype of $d$-wave and $p$-wave superconductor junctions. 
In $d_{x^{2}-y^{2}}$-wave junctions, 
$\sigma_{\rm T}(eV)$ has a peak splitting
[see curve $b$ and $c$ in Fig.~\ref{fig:01}(a)]
in the presence of magnetic field $H$ \cite{Fogel,Tana02} 
since the transparency of the 
junction is sufficiently small.
%%%%%%%%%%%%%%%%%%%%%%%%%%%%%%%%%
%p-wave 
%%%%%%%%%%%%%%%%%%%%%%%%%%%%%
On the other hand, in the $p_{x}$-wave junctions, 
there is no ZBCP splitting with the increase of the 
magnitude of $H$, where only the height of the ZBCP is reduced  
and the width becomes broad [Fig.\ref{fig:01}(b)]. 
Thus, the response of ZBCP to the magnetic field is quite
different from that in the $d_{x^{2} -y^{2}}$-wave case. 
\par
%%%%%%%%%%%%%%%%%%%%%%%%%%%%%%
% Angle resolved analysis
%%%%%%%%%%%%%%%%%%%%%%%%%%%%%%%%%%%
%--- figure 2 ---
\begin{figure}[htb]
\begin{center}
%\epsfxsize=8cm
%\epsfysize=14cm
%\centerline{\epsfbox{fig2x.eps}}
\includegraphics[scale=.5]{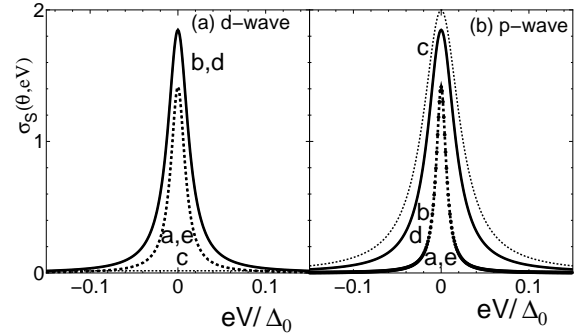}
\caption{
Angle resolved tunneling conductance 
in the superconducting state $\sigma_{\rm S}(\theta,eV)$ 
for $H=0$ with $\lambda_{0}d_{i}=3$ and 
$\kappa=0.5$. 
(a) $d_{x^{2}-y^{2}}$-wave and 
and 
(b) $p_{x}$-wave. 
a: $\theta=-\pi/4$,
b:$\theta=-\pi/8$, 
c: $\theta=0$,
d: $\theta=\pi/8$ and 
e: $\theta=\pi/4$. 
\label{fig:02}}
\end{center}
\end{figure}
%--- figure 3 ---
\begin{figure}[htb]
\begin{center}
%\epsfxsize=8cm
%\epsfysize=14cm
%\centerline{\epsfbox{fig3x.eps}}
\includegraphics[scale=.5]{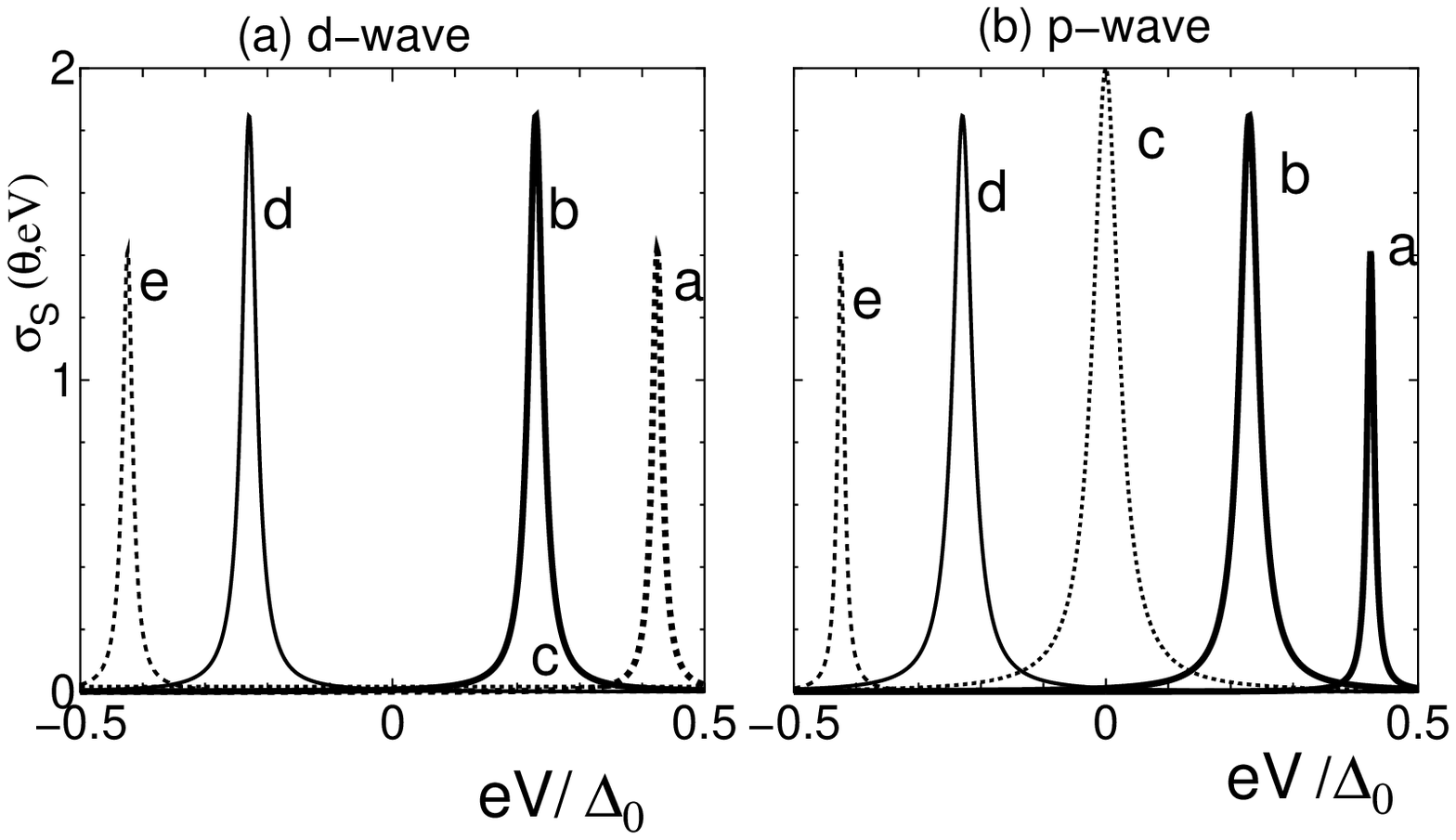}
\caption{
Angle resolved tunneling conductance 
in the superconducting state $\sigma_{\rm S}(\theta,eV)$ 
for $H=0.2H_{0}$ with $\lambda_{0}d_{i}=3$ and 
$\kappa=0.5$. 
(a) $d_{x^{2}-y^{2}}$-wave and 
(b) $p_{x}$-wave. 
a: $\theta=-\pi/4$,
b:$\theta=-\pi/8$, 
c: $\theta=0$,
d: $\theta=\pi/8$ and 
e: $\theta=\pi/4$. 
\label{fig:03}}
\end{center}
\end{figure}
%---------------------------------------------------------------
In order to understand this difference, 
we look into 
$\sigma_{\rm S}(\theta,eV)$, $i.e.$,
the angle resolved conductance in the 
superconducting state. 
For $H=0$, the height of ZBCP 
$\sigma_{\rm S}(\theta,0)$ 
is given as  
$\sigma_{S}(\theta,0)=2\cos\theta$ for 
$f(\theta) \neq 0$ 
independent of $\sigma_{\rm N}(\theta)$
due to the formation of ZES and 
$\sigma_{\rm S}(\theta,0)=\sigma_{N}(\theta)$ for 
$f(\theta) = 0$, respectively 
\cite{Tanaka95}. 
In the absence of the magnetic field, 
$\sigma_{\rm S}(\theta,eV)$ always 
has ZBCP except for 
special cases, $i.e.$, $\theta=0$ for the 
$d_{x^{2}-y^{2}}$-wave junction 
and $\theta= \pm \pi/2$ for both junctions. %$p$-wave junction.
For the $d_{x^{2}-y^{2}}$-wave junction, the width of the ZBCP $W$ takes its 
maximum value for some oblique injection angle $\theta_{m}$,   
while for $p_{x}$-wave, both $W$ and 
$\sigma_{\rm S}(\theta,0)$  becomes largest at $\theta=0$.
\par
By applying $H$, the positions of peak shift from zero. 
In the present case, since positive $H$ is chosen, 
quasiparticle energy $E$ increases (decreases) 
for $\theta >0$ ($\theta<0$), so that the 
peak position of $\sigma_{\rm S}(\theta,eV)$ 
becomes located in the negative (positive) voltage region. 
%%%%%%%%%%
However, a remarkable feature is that the 
ZBCP for perpendicular injection, $i.e.$ 
$\sigma_{\rm S}(0,eV)$, is not changed at all by $H$. 
This is because $v_{{\rm F}y}=0$ at $\theta=0$, so that 
there is no Doppler shift for perpendicular injection.
For the $p_{x}$-wave case, perpendicular injection 
contributes most dominantly to the low bias behavior 
of the integrated 
normalized tunneling conductance $\sigma_{\rm T}(eV)$, 
and consequently the ZBCP is robust against $H$. 
%
%%%%%%%%%%%%%%%%%%%%%%%%%%%%%%%%%%
% general discussion 
%%%%%%%%%%%%%%%%%%%%%%%%%%%%%%%%%%
%
In the above, $v_{{\rm F}y}=0$ at $\theta=0$ is
a consequence of assuming a cylindrical Fermi surface.
In fact, the absence of Doppler shift 
for perpendicular injection can be shown in more generalized 
cases with $v_{{\rm F}y}(k_{x},k_{y}=0) \neq 0$. 
Namely, there is generally an inversion symmetry of the 
crystal, $\epsilon(\mib{k})=\epsilon(-\mib{k})$,  
so the injected and reflected quasiparticle states 
have different sign of $v_{{\rm F}y}$
($v_{{\rm F}y}(k_{x},0)= -v_{{\rm F}y}(-k_{x},0)$) 
for perpendicular injection. 
%
%We can interpret that $v_{Fy}=0$ corresponds 
%to  a quite special case. 
%
In order to look into this general situation, 
we consider a phenomenological analogue of
$\sigma_{\rm R}(\theta,E)$ as 
%
%-------------Surface Density of States----------------------------------------
\begin{eqnarray}
\rho(E)
=\frac{1+ \bar{\sigma} |\Gamma_{+}|^{2}
+[\bar{\sigma} - 1]
|\Gamma_{+}|^{2}|\Gamma_{-}|^{2}}
{|1+[\bar{ \sigma} - 1]
\Gamma_{+}\Gamma_{-}|^{2}}.
\end{eqnarray}
Here we take $\bar{\sigma}=0.1$ as a typical value for a  
low transparent barrier, and calculate $\rho(E)$ for two cases, $i.e.$, 
i) $v_{{\rm F}y}(k_{x},0)=-v_{{\rm F}y}(-k_{x},0)$  
$i.e.$, $g_{+} = -g_{-}$, 
and 
ii) $v_{{\rm F}y}(k_{x},k_{y})=v_{{\rm F}y}(-k_{x},k_{y})$ 
$i.e.$, $g_{+} = g_{-}$.  
%
%--- figure 4 ---
\begin{figure}[htb]
\begin{center}
%\epsfxsize=8cm
%\epsfysize=14cm
%\centerline{\epsfbox{fig5x.eps}}
\includegraphics[scale=.5]{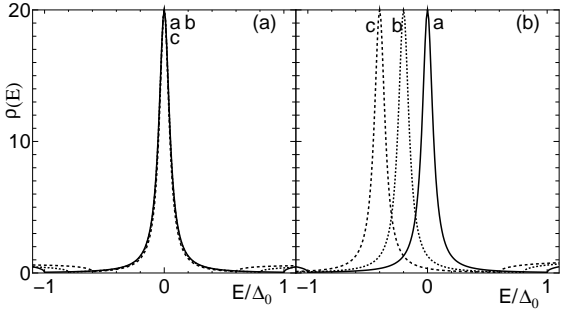}
\caption{
Local density of state 
$\rho(E)$ with 
(a) $v_{{\rm F}y}(k_{x})=-v_{{\rm F}y}(-k_{x})$
and 
(b) $v_{{\rm F}y}(k_{x})=v_{{\rm F}y}(-k_{x})$. 
a: $H=0$,
b: $H=0.2H_{0}$,
c: $H=0.4H_{0}$.
\label{fig:04}}
\end{center}
\end{figure}
As seen from Fig.~\ref{fig:04}(a),
there is no Doppler shift of 
ZBCP with the increase of $H$ for case (i).  
The absence of the shift 
is quite different from that of 
the conventional case (ii) [Fig.~\ref{fig:04}(b)], 
where the magnitude of the shift of the peak position
is proportional to the magnetic field \cite{Fogel}.
The absence of Doppler shift originates from the 
cancellation of the additional phase in the product of 
$\Gamma_{+}\Gamma_{-}$ due to  the different sign of 
$\tilde{E}_{+}$ and $\tilde{E}_{-}$ at $E=0$. 
The cancellation of the additional phase shift 
of the wave function is a quite novel feature 
originating from the difference
in the sign of $v_{{\rm F}y}$ 
between reflected and injected quasiparticle.
Although the present argument
for $v_{{\rm F}y}(k_x,0)=-v_{{\rm F}y}(-k_x,0)\neq 0$ 
is phenomenological, recently,
we have shown for a tight binding model 
of a possibly triplet superconductor $\mbox{(TMTSF)}_{2}X$
that zero-energy peak in the surface density of 
states does not split in the presence of a magnetic field, 
not only for $p$-wave pairing, but also for a triplet $f$-wave pairing.
\cite{TAU2002}. \par
\begin{figure}[htb]
\begin{center}
%\epsfxsize=8cm
%\epsfysize=14cm
%\centerline{\epsfbox{fig06.eps}}
\includegraphics[scale=.5]{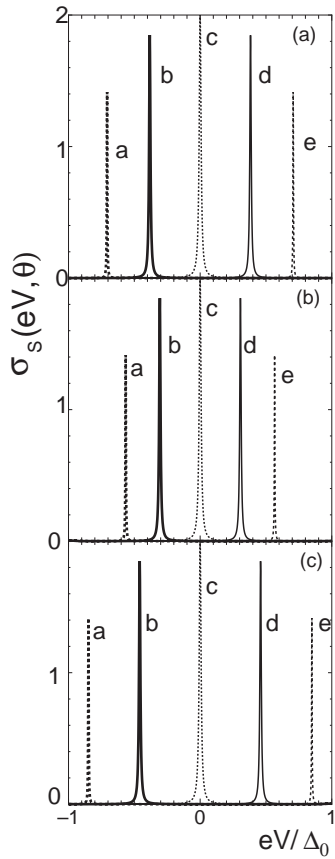}
\caption{
Angle resolved tunneling conductance 
in the superconducting state $\sigma_{\rm S}(\theta,eV)$ 
in $p_{x}$+i$p_{y}$-wave state 
with $\lambda_{0}d_{i}=3$ and $\kappa=0.5$. 
(a) $H=0$, 
(b) $H=0.2H_{0}$ and  
(c) $H=-0.2H_{0}$. 
a: $\theta=-\pi/4$, 
b: $\theta=0$, and 
c: $\theta=\pi/4$. 
\label{fig:06}}
\end{center}
\end{figure}

Finally, we now look into the chiral $p$-wave ($p_x$+i$p_y$-wave) case, 
where $f(\theta)$ is given as $f(\theta) = \exp({\rm i} \theta)$. 
This is a pairing symmetry possibly realized in Sr$_{2}$RuO$_{4}$.
%%%%%%%%%%%%%%%%%%%%%%%
% chiral super
%%%%%%%%%%%%%%%%%%%%%%%
%---------------------------------------------------------------
%\newpage
\begin{figure}[htb]
\begin{center}
%\epsfxsize=8cm
%\epsfysize=15cm
%\centerline{\epsfbox{fig5x.eps}}
\includegraphics[scale=.5]{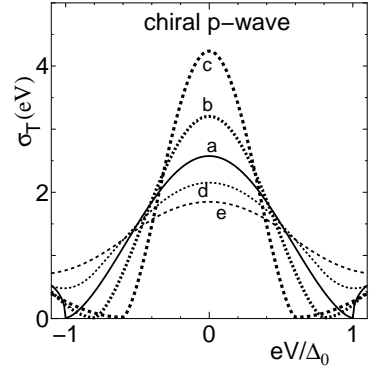}
\caption{
The normalized tunneling conductance $\sigma_{\rm T}(eV)$
in the $p_{x}$ + i$p_{y}$-wave state. 
$\lambda_{0}d_{i}=3$ and 
$\kappa=0.5$. 
a: $H=0$, b: $H=0.2H_{0}$, c: $H=0.4H_{0}$
d: $H=-0.2H_{0}$ and e: $H=-0.4H_{0}$. 
\label{fig:05}}
\end{center}
\end{figure}
As seen in Fig.~\ref{fig:05}, ZBCP again does not split in the presence of a 
magnetic field, while its height is enhanced for 
positive $H$ ($H$ in the $+z$ direction; curves $b$ and $c$), 
but is reduced for negative $H$ ($H$ in the $-z$ direction; 
curves $d$ and $e$). 
Such an asymmetric $H$ dependence of ZBCP 
around $H=0$ does not appear 
in $d_{x^{2}-y^{2}}$-wave and $p_{x}$-wave cases.  
In order to understand this asymmetric feature in detail,
we also plot in Fig.~\ref{fig:06}
the angle resolved tunneling conductance 
as in Figs.~\ref{fig:02} and \ref{fig:03}. 
%%%%%%%%%%%%%%%%%%%%%%%%%%%%%%%%%%%%%
%  Angle resolved conductance
%%%%%%%%%%%%%%%%%%%%%%%%%%%%%%%%%%%%%%%5
%--- figure 6 ---
%-------------------------------------
As seen in Fig.~\ref{fig:06},
the peak position of  
$\sigma_{\rm S}(\theta,eV)$ has a strong
$\theta$ dependence. 
%
%If we denote the angle dependent peak position as 
%$E_{p}(\theta)$, 
%$-E_{p}(-\pi/4)=E_{p}(\pi/4) \sim 0.71\Delta_{0}$ 
%$-E_{p}(-\pi/8)=E_{p}(\pi/8) \sim 0.38\Delta_{0}$ 
%and $E_{p}(0)=0$. 
%
Namely, the $p_x$+i$p_y$-wave pairing induces 
a broken time reversal symmetry state (BTRSS),
so that the peak positions for $\theta\neq 0$
is no longer located at $E=0$ \cite{Kashi95b}
even in the absence of a magnetic field. 
Now, in the presence of a positive magnetic field, 
all the peaks for $\theta\neq 0$ are shifted {\it toward} $E=0$,
[Fig.~\ref{fig:06}(b)] while they are shifted
{\it away from} $E=0$ for a negative magnetic field
[Fig.~\ref{fig:06}(c)].
%
%$E_{p}(\theta)$ becomes  
%$-E_{p}(-\pi/4)=E_{p}(\pi/4) \sim 0.57\Delta_{0}$, 
%$-E_{p}(-\pi/8)=E_{p}(\pi/8) \sim 0.31\Delta_{0}$ 
%and $E_{p}(0)=0$ for $H=0.2H_{0}$,  
%and 
%$-E_{p}(-\pi/4)=E_{p}(\pi/4) \sim 0.85\Delta_{0}$, 
%$-E_{p}(-\pi/8)=E_{p}(\pi/8) \sim 0.46\Delta_{0}$ 
%and $E_{p}(0)=0$ for $H=-0.2H_{0}$, respectively.    
%
%The positive $H$ enhances the magnitude of $E_{p}$ 
%while negative one does not. 
%
This unique feature can be interpreted as follows.  
In $p_{x}$+i$p_{y}$ state, due to the formation of 
BTRSS, there is a surface current \cite{Matsumoto}
which flows parallel to the interface 
and the spontaneous magnetic field 
even without applied magnetic field. 
If the direction of the  applied magnetic field $H$
is the same (opposite) as that of spontaneous field, 
the effective magnetic field is 
enhanced (reduced)  for $H<0$ ($H>0$). 
Then the resulting magnitude of ZBCP in total conductance 
$\sigma_{\rm T}(eV)$ 
is reduced (enhanced) for negative (positive) $H$. 
%==================================================
%\section{conclusion}
%==================================================
%
In summary, we have calculated 
tunneling conductance  in 
normal metal / unconventional superconductor
junctions, 
where $d_{x^2-y^2}$-wave,
$p_{x}$-wave and $p_{x}$+i$p_{y}$-wave pairings
have been chosen as a prototype. 
We focused on the influence of the applied 
magnetic field on ZBCP. 
For $p$-wave cases, 
ZBCP does not split into two by magnetic field  
since the most dominant contribution to tunneling conductance 
at zero-energy originates from perpendicular injection,  
where Doppler shift does not occur.
The absence of Doppler shift has been shown in general 
situations, where the component of the Fermi velocity 
parallel to the interface has the same magnitude but 
different signs between 
injected and reflected quasiparticle states.
As for $p_{x}$+i$p_{y}$-wave superconductor junctions, 
we have shown that ZBCP does not split, 
while its height is sensitive to the direction 
of the applied magnetic field, which is a consequence of  
a broken time reversal symmetry.
This behavior should be observed in the 
tunneling experiments of Sr$_{2}$RuO$_{4}$,
if it is actually a chiral $p$-wave superconductor.
\par
%==================================================
The authors acknowledge H. Aoki and R. Arita for 
discussions on $\mbox{(TMTSF)}_{2}X$
This work was supported by
the Core Research for
Evolutional Science and Technology (CREST)
of the Japan Science
and Technology Corporation (JST).
%
%The computational aspect of this work has been performed
%at the facilities of the Supercomputer Center,
%Institute for Solid State Physics,
%University of Tokyo and the Computer Center.
\par
%====Reference===================================
%
%-----------------------------------------------

%
\end{document}